**Joint Optimization of Pattern, Headway, and Fleet Size of Multiple Urban Transit Lines with Perceived Headway Consideration and Passenger Flow Allocation**

**Max T.M. Ng**
Graduate Research Assistant
Northwestern University Transportation Center
600 Foster Street
Evanston, IL 60208, USA
Email: maxng@u.northwestern.edu

**Draco Tong**
Graduate Research Assistant
Northwestern University Transportation Center
600 Foster Street
Evanston, IL 60208, USA
Email: draco.tong@northwestern.edu

**Hani S. Mahmassani**[†]
William A. Patterson Distinguished Chair in Transportation
Director, Transportation Center
Northwestern University
600 Foster Street
Evanston, IL 60208, USA

**Ömer Verbas**
Technical Lead for Network Modeling and Simulation
Transportation Systems & Mobility Group, Vehicle & Mobility Systems Department
Argonne National Laboratory
Lemont, IL 60439, USA
Email: omer@anl.gov

**Taner Cokyasar**
Consultant
The Transportation and Power Systems Division
Argonne National Laboratory
Lemont, IL 60439, USA
Email: tcokyasar@anl.gov

[†] Deceased

## ABSTRACT

This study addresses the urban transit pattern design problem, optimizing stop sequences, headways, and fleet sizes across multiple routes and periods simultaneously to minimize user costs (composed of riding, waiting, and transfer times) under operational constraints (e.g., vehicle capacity and fleet size). A destination-labeled multi-commodity network flow (MCNF) formulation is developed to solve the problem at a large scale more efficiently compared to the previous literature. The model allows for flexible pattern options without relying on pre-defined candidate sets and simultaneously considers multiple operational strategies such as express/local services, short-turning, and deadheading. It evaluates perceived headways of joint patterns for passengers, assigns passenger flows to each pattern accordingly, and allows transfers across patterns in different directions. The mixed-integer linear programming (MILP) model is demonstrated with a city-sized network of metro lines in Chicago, IL, USA, achieving near-optimal solutions in hours. The total weighted journey times are reduced by 0.61% and 5.76% under single-route and multi-period multi-route scenarios respectively. The model provides transit agencies with an efficient tool for comprehensive service design and resource allocation, improving service quality and resource utilization without additional operational costs.

# 1 INTRODUCTION

## 1.1 Motivation

The efficiency of transit systems is crucial for serving passengers within a tight budget. Many systems have seen significant changes in demand patterns since their original design decades ago, with more pronounced shifts following the COVID-19 pandemic, which has altered commuting behaviors and increased the demand for flexible, efficient transit solutions (Tahlyan et al., 2022). For example, there has been discussion of converting commuter rail services to regional rail models and consolidating transit agencies in North American cities (Cano, 2024; Freishtat, 2024). These changes, coinciding with the future potential redesign of transit networks with shared autonomous mobility services (Ng et al., 2024), necessitate a redesign of patterns to maintain efficiency within constrained operational budgets.

Train service design problem has been well studied, considering a range of operational strategies to improve efficiency and services, such as route splitting (Wilson and González, 1982), zonal express/local services (Furth, 1986), short-turning and deadheading (Ceder, 1989; Furth, 1987), and skip-stop strategies (Peftitsi et al., 2023). More recent literature has formulated the problem with mathematical programming and solved different variants of the problems considering some operational strategies and small problem instances (Feng et al., 2024; Leiva et al., 2010; Ulusoy et al., 2010; Wang et al., 2018). Solution approaches that simultaneously consider multiple pattern designs, headways, and fleet sizes across routes can better support transit agencies in optimizing their services given a budget constraint.

## 1.2 Problem Description

This study addresses the service design problem for urban rail transit lines, optimizing stop sequences and headways of each pattern simultaneously to minimize user costs (composed of riding, waiting, and transfer times) under operational constraints (e.g., vehicle capacity and fleet size). Unlike line[1] planning problems that determine patterns across multiple lines in a network (Schöbel, 2012), this study assumes pre-determined routes. This approach aligns more closely with train service route design and passenger flow allocation problems as examined by Feng et al. (2024).

The train service design problem is combinatorial as the solution space explodes with possible service pattern combinations. It is also non-linear due to the relationship between waiting time and fleet requirements when setting headway. These limit previous solution approaches in solving larger or city-scale problems.

This study develops a model that allows for flexible pattern options without relying on pre-defined candidate sets, and simultaneously considers multiple operational strategies, such as route splitting and express/local services. It also models bi-directional operations, including short-turning and deadheading, assuming continuous operations in each of the multiple periods (e.g., peak and off-peak hours). Demand is exogenous and inelastic, input into the model as origin-to-destination (O-D), stop-by-stop data. Headway is determined from a discrete set that overcomes the nonlinearity with fleet requirement.[2] Passenger flow is constrained by the capacity provided by the headway. We also allow transfers across patterns in different directions.

[1] The terms “line” and “route” are used interchangeably in this paper.

Previous research investigated perceived headway when multiple patterns serve the same O-D pairs (Figure 1) (Verbas and Mahmassani, 2013). Passengers can take any minimum-cost path (Nguyen and Pallottino, 1989; Spiess and Florian, 1989; Verbas and Mahmassani, 2015a), considering weighted waiting, riding, and transfer times, subject to capacity constraints. Additionally, the assumption that passengers board the first available vehicle leads to the frequency share rule (Desaulniers and Hickman, 2007) where passenger numbers for the same O-D pair are distributed among different patterns based on their frequencies. This study explicitly considers the perceived headway of joint patterns and assigns passenger flows to each pattern accordingly.

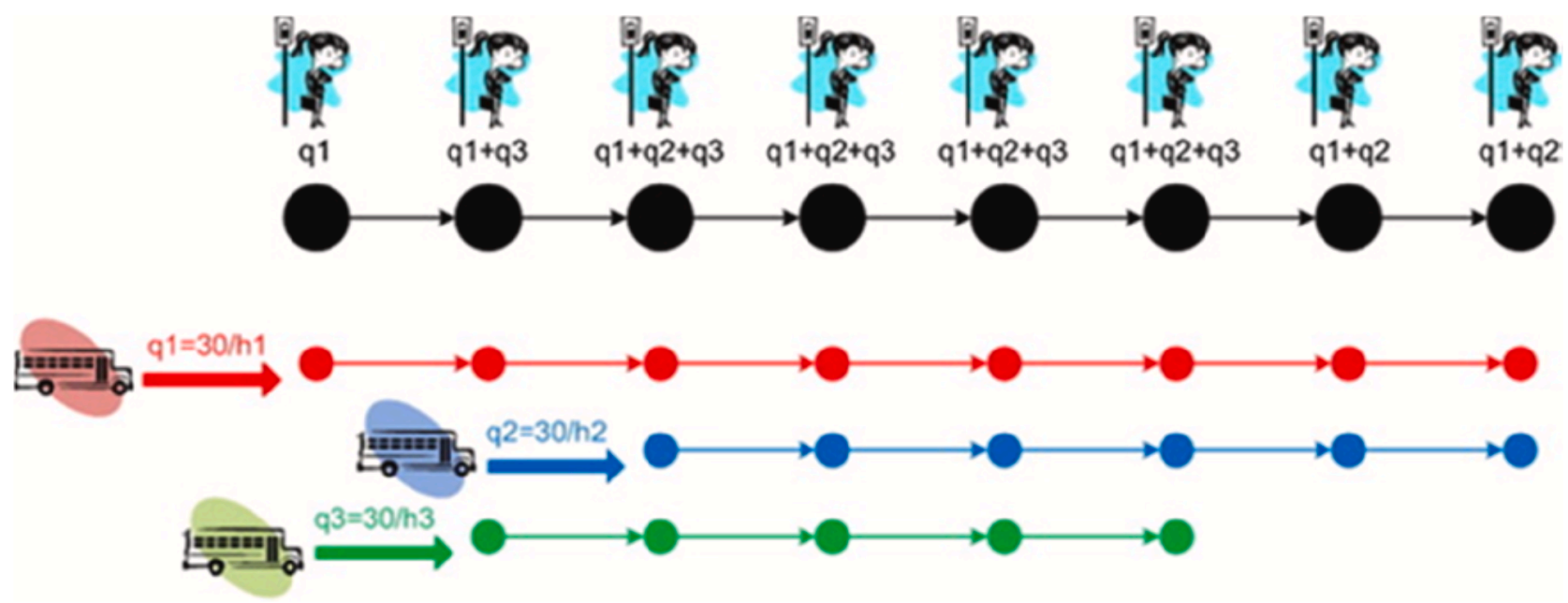

**Figure 1. Illustration of transit service patterns along a transit route. (Verbas and Mahmassani, 2013)**

We further expand the model to optimize patterns and headways of multiple routes and periods (e.g., peak and off-peak hours), while assigning trains[3] to each route from a pool. This broader approach captures the interconnected nature of urban transit systems, ensuring that fleet resources are allocated efficiently.

## 1.3 Contributions

This study presents theoretical and methodological contributions. Theoretically, this study formulates the transit service pattern design problem that jointly optimizes stop patterns, headways, and fleet sizes for metro line routes. This formulation is more general compared to previous works on train service design, as it allows multiple service strategies, considers perceived headways and flow assignment of multiple patterns, and includes transfers across patterns. Methodologically, we develop a destination-labeled multi-commodity network flow (MCNF) formulation, leading to efficient mixed-integer linear programming (MILP) solutions that address the problem at a large scale (more stops, multiple routes, and multiple periods) faster compared to previous literature. Binary variables assign stop sequences and headways to patterns, and perceived headways to O-D pairs, while continuous variables assist in evaluating flows and costs. The methodology is demonstrated with a case study of metro lines in Chicago, USA.

[2] While this may seem a limiting factor for choices of headway, it is close to real-world operations where train schedules are set at regular times (e.g., in minutes) instead of arbitrary decimals.
[3] Train or fleet size is chosen as one of the optimization units instead of train cars for scenarios where train drivers or train sets are the constraints. This is particularly relevant in systems where labor shortage poses a more critical constraint. Nevertheless, the formulation can be readily adapted to consider train car assignment as well.

The remainder of the paper is structured as follows. Section 2 reviews the literature on the transit service design problem and headway and fleet size setting. Section 3 describes the MCNF model for single-route and multi-route formulations. Section 4 introduces the experiment setting of both single-route and multi-route problems, followed by the result discussion in Section 5. Section 6 concludes the paper with limitations and future research directions.

# 2 BACKGROUND

## 2.1 Transit Service Design Problem

The general transit network design problem has been extensively reviewed by Durán-Micco and Vansteenwegen (2022) and Farahani et al. (2013). While focusing on bus lines, Ibarra-Rojas et al. (2015) provided a comprehensive review of the literature on service design problems, highlighting the uncovered research area of design such as short-turning and express lines.

Several rule-based and heuristic approaches have been proposed to optimize the deployment of various operating strategies in transit lines. These include route splitting (Wilson and González, 1982), zonal express/local services (Furth, 1986), short-turning and deadheading (Ceder, 1989; Furth, 1987), and skip-stop strategies (Peftitsi et al., 2023). Gkiotsalitis et al. (2019) studied short-turning and interlining with optimal frequency & fleet size. Additionally, Baaj and Mahmassani (1991, 1995) developed general route generation heuristics.

More recent and growing literature has focused on mathematical programming approaches. Leiva et al. (2010) enumerated paths to design limited stop services. Ulusoy et al. (2010) considered all-stop, short-turning, and express options with transfers. Yang et al. (2017) assigned passengers with a frequency logit model in designing short-turning patterns. Wang et al. (2018) developed a generic approach but with a limited number of stops. Chen et al. (2018) incorporated local service and short-turning in the greater transit network design problem, and Li et al. (2021) optimized patterns and frequencies with short-turning and train compositions. Feng et al. (2024) designed train service routes (express/local) with passenger flow allocation.

This study distinguishes itself by not relying on pre-defined candidate pattern sets, allowing for more flexible pattern options. It offers a more holistic consideration of multiple service strategies simultaneously and addresses problems at a large scale (current study considering lines with at most 40 stops as reviewed by Feng et al. (2024)). We jointly optimize pattern, headway, and fleet size while taking into account perceived headway and performing trip assignments.

## 2.2 Transit Headway and Fleet Size Setting

While transit frequency and fleet size were usually considered jointly within the transit network design problem (e.g., Pinto et al., 2020), a considerable portion of the work focused specifically on optimizing headway and fleet assignment across routes assuming a designed network (e.g.,

Verbas and Mahmassani, 2013, 2015b). Verbas et al. (2015) further optimized bus frequency with elastic demand. Martínez et al. (2014) set frequency with flow assignment, utilizing MILP for small instances, and metaheuristics for larger problems. Other problem types include frequency and pricing setting (Bertsimas et al., 2020), frequency and schedule setting with bounded stochastic user equilibrium (Jiang et al., 2022), and frequency in line planning problem with piecewise linear approximation of waiting time (Zhou et al., 2021).

These studies collectively provide a foundation for understanding the complexity and interdependencies involved in the joint problem of transit service design and headway and fleet size setting. We build on this foundation by developing more integrated and scalable solutions to optimize transit service design.

# 3 MODEL

## 3.1 Nomenclature

The notation is listed in Table 1. Sets are denoted as capital scripted characters (e.g., $\mathcal{S}_r$ and $\mathcal{P}_r$ ), constants as Greek or capital Roman characters (e.g., $\gamma^a$ and $T^{\lambda}_{ij}$), and indices and variables as small Roman characters (e.g., $d$ and $f^{\alpha}_{tdicp}$). Superscripts are qualifiers and subscripts are indices. The cardinality of set $\mathcal{S}_r$ is denoted with $|\mathcal{S}_r|$.

**Table 1. Notation.**

| ***Sets and indices*** | |
|---|---|
| $\mathcal{C}_r$ | Set of headway combinations of route $r \in \mathcal{R}$, indexed by $c \in \mathcal{C}_r$ |
| $\mathcal{D}_r$ | Set of origins/destinations of route $r \in \mathcal{R}$, indexed by $o \in \mathcal{D}_r$ and $d \in \mathcal{D}_r$, respectively |
| $\mathcal{H}_r$ | Ordered set of headway choices of route $r \in \mathcal{R}$, indexed by $h \in \mathcal{H}_r$ |
| $\mathcal{P}_r$ | Set of patterns of route $r \in \mathcal{R}$, indexed by $p \in \mathcal{P}_r$ |
| $\mathcal{R}$ | Set of routes, indexed by $r \in \mathcal{R}$ |
| $\mathcal{S}_r$ | Set of stops of route $r \in \mathcal{R}$, indexed by $i, j, k \in \mathcal{S}_r$ |
| $\mathcal{T}$ | Set of time periods, indexed by $t \in \mathcal{T}$ |
| $h^{\gamma}_{cp}$ | Headway index taken by pattern $p \in \mathcal{P}_r$ under combination $c \in \mathcal{C}_r$, where $h^{\gamma}_{cp} = 0$ indicates that the service is unavailable |
| ***Parameters*** | |
| $B_r$ | Vehicle capacity of route $r \in \mathcal{R}$ |
| $D_t$ | Duration of period $t \in \mathcal{T}$ |
| $E_{tod}$ | Demand between origin $o \in \mathcal{D}_r$ and destination $d \in \mathcal{D}_r$ of period $t \in \mathcal{T}$ |
| $M$ | A sufficiently large number (big-M) |
| $N^{v}$ | Number of vehicles available |
| $N^{\tau}$ | Number of vehicle-hours available |
| $T^{\chi}$ | Transfer cost |
| $T^{\lambda}_{ij}$ | Travel time between stops $i \in \mathcal{S}_r$ and $j \in \mathcal{S}_r$ |
| $T^{\omega}_{c}$ | Perceived headway for combination $c \in \mathcal{C}_r$ |
| $T^{\gamma}_{h}$ | Headway of index $h \in \mathcal{H}_r$ |
| $\gamma^{\chi}$ | Transfer cost coefficient relative to travel time |

| $\gamma^{\omega}$ | Waiting cost coefficient relative to travel time |
|---|---|
| ***MILP decision variables*** | |
| $f^{\alpha}_{tdicp}$ | Boarding flows at stop $i \in \mathcal{S}_r$ to destination $d \in \mathcal{D}_r$ via headway combination $c \in \mathcal{C}$ to trains of pattern $p \in \mathcal{P}_r$ in period $t \in \mathcal{T}$ |
| $f^{\beta}_{tip}$ | Exit flows at stop $i \in \mathcal{S}_r$ from trains of pattern $p \in \mathcal{P}_r$ in period $t \in \mathcal{T}$ |
| $f^{\chi}_{tdijpc}$ | Transfer flows at stop $i \in \mathcal{S}_r$ (to stop $j \in \{i, \tilde{\imath}\}$) from trains of pattern $p \in \mathcal{P}_r$ to headway combination $c \in \mathcal{C}_r$, with destination $d \in \mathcal{D}_r$ in period $t \in \mathcal{T}$ |
| $f^{\lambda}_{tdpij}$ | Inter-stop flows from stop $i \in \mathcal{S}_r$ to stop $j \in \mathcal{S}_r$ on trains of pattern $p \in \mathcal{P}_r$ with destination $d \in \mathcal{D}_r$ in period $t \in \mathcal{T}$ |
| $f^{\omega}_{tdic}$ | Entry flows at stop $i \in \mathcal{S}_r$ to destination $d \in \mathcal{D}_r$ to headway combination $c \in \mathcal{C}_r$ in period $t \in \mathcal{T}$ |
| $x_{tpij}$ | Binary variable indicating if stop $j \in \mathcal{S}_r$ is the next stop of $i \in \mathcal{S}_r$ in pattern $p \in \mathcal{P}_r$ in period $t \in \mathcal{T}$ |
| $x^{\eta}_{tpijh}$ | Binary variable indicating if stop $j \in \mathcal{S}_r$ is the next stop of $i \in \mathcal{S}_r$ in pattern $p \in \mathcal{P}_r$ *and* the headway of pattern $p$ is $h \in \mathcal{H}_r$ in period $t \in \mathcal{T}$ |
| $y_{tph}$ | Binary variable indicating if the headway of pattern $p$ is $h \in \mathcal{H}_r$ in period $t \in \mathcal{T}$ |
| $z_{tidc}$ | Binary variable indicating if passengers at stop $i \in \mathcal{S}_r$ with destination $d \in \mathcal{D}_r$ is assigned the headway combination $c \in \mathcal{C}_r$ in period $t \in \mathcal{T}$ |
| ***Additional decision variable in multi-route joint optimization*** | |
| $n^{v}_{rt}$ | Fleet size of route $r \in \mathcal{R}$ in period $t \in \mathcal{T}$ |

Boarding in opposite directions at the same physical stop is represented by $i$ and $\tilde{\imath}$. Formally, $\tilde{d} = |\mathcal{S}_r| - d - 1$ and $[\widetilde{d_1, d_2}] = [\widetilde{d_2}, \widetilde{d_1}]$.[4] The same notation applies to a set, e.g., $\widetilde{\mathcal{D}_r}$, so $\mathcal{D}_r \cup \widetilde{\mathcal{D}_r} = \mathcal{S}_r$.

We optimize the patterns and headways jointly for multiple routes $\mathcal{R}$ and periods $\mathcal{T}$. The simpler cases of single route or single period would reduce the optimization problem by shrinking the corresponding dimensions of variables (i.e., $r$ and $t$).

### 3.2 Perceived Headway

We assume that passengers take any minimum cost path (considering weighted waiting, riding, and transfer times and subject to capacity constraints). However, passengers of certain O-D pairs may be provided with more than one pattern to reach their destinations, or in other words, they can travel on trains of any pattern with similar journey times. We further assume that passengers take the first vehicle passing by in such cases. This leads to a perceived headway lower than any relevant pattern headway, or mathematically $1/(\sum_{h \in \mathcal{H}_r} 1/T^{\gamma}_{h})$ , where $T^{\gamma}_{h}$ is the headway of index $h \in \mathcal{H}_r$[5], as the frequency share rule commonly adopted in the literature (Desaulniers and Hickman, 2007; Nguyen and Pallottino, 1989).

To evaluate the perceived headway linearly, each unique combination $c \in \mathcal{C}_r$ of headway $h \in \mathcal{H}_r$ and pattern $p \in \mathcal{P}_r$ for an O-D pair is pre-computed and iterated over. This allows the model to consider the smaller perceived headway resulting from the operation of more than one pattern. Eq.(1) formally defines the set of combination $\mathcal{C}_r$ with $h^{\gamma}_{cp}$ as the headway index taken by pattern $p$ under each combination $c$, where $h^{\gamma}_{cp} = 0$ indicates that pattern $p$ is out of service.

[4] Stop numbering starts with 0.

[5] $\mathcal{H}_r$ consists of the headway choices and $h = 0$ to indicate not-in-service.

$$\mathcal{C}_r = \mathcal{H}_r^{\mathcal{P}_r} \backslash \{(0,0,\dots,0)\} = \left\{ \left( h^{\gamma}_{c,1}, h^{\gamma}_{c,2}, \dots, h^{\gamma}_{c,|\mathcal{P}_r|} \right) : h^{\gamma}_{cp} \in \mathcal{H}_r, \exists h^{\gamma}_{cp} \neq 0 \right\} \quad \forall r \in \mathcal{R} \tag{1}$$

For example, $c = (0,1)$ means first pattern is not available and second pattern offers first headway (e.g., 5-minute).

The set of combinations with two patterns and two headways (0 indicating no service; 1 and 2 indicating the first and second headway choice) is given by: $\mathcal{C}_r = \{(0,1),(0,2),(1,0),(1,1),(1,2),(2,0),(2,1),(2,2)\}$. Therefore, the number of possible combinations is $|\mathcal{C}_r| = |\mathcal{H}_r|^{|\mathcal{P}_r|} - 1$, accounting for all possible headway combinations (minus one for the all out-of-service pattern case). For example, $|\mathcal{C}_r| = 8$ for two patterns and two non-zero headways, and $|\mathcal{C}_r| = 63$ for three patterns and three non-zero headways.

Each O-D pair experiences a specific headway-pattern combination. The perceived headway of a combination $c \in \mathcal{C}_r$ under a joint frequency served by multiple patterns is determined by $T^{\omega}_c$ in Eq.(2).

$$T^{\omega}_c = \left( \sum_{p \in \mathcal{P}_r | h^{\gamma}_{cp} \neq 0} \left( T^{\gamma}_{h^{\gamma}_{cp}} \right)^{-1} \right)^{-1} \quad \forall c \in \mathcal{C}_r, \forall r \in \mathcal{R} \tag{2}$$

By the frequency share rule, among passengers boarding at stop $i \in \mathcal{S}_r$ to go to destination $d \in \mathcal{D}_r$, the portion using pattern $p_1$ (with flow $f^{\alpha}_{tdicp_1}$) is also the portion of its frequency (inverse of headway) to the total frequency, i.e., Eq.(3). This leads to the ratios between the flows of any two patterns in Eq.(4).

$$\frac{f^{\alpha}_{tdicp_1}}{\sum_{p \in \mathcal{P}_r} f^{\alpha}_{tdicp}} = \frac{\left( T^{\gamma}_{h_1} \right)^{-1}}{\sum_{p \in \mathcal{P}_r} \left( T^{\gamma}_{h^{\gamma}_{cp}} \right)^{-1}} \quad h_1 = h^{\gamma}_{cp_1}, \forall i \in \mathcal{S}_r, d \in \mathcal{D}_r, p_1 \in \mathcal{P}_r, c \in \mathcal{C}_r, \forall r \in \mathcal{R}, t \in \mathcal{T} \tag{3}$$

$$\frac{f^{\alpha}_{tdicp_1}}{f^{\alpha}_{tdicp_2}} = \frac{T^{\gamma}_{h_2}}{T^{\gamma}_{h_1}} \quad h_1 = h^{\gamma}_{cp_1}, h_2 = h^{\gamma}_{cp_2}, \forall i \in \mathcal{S}_r, d \in \mathcal{D}_r, p_1, p_2 \in \mathcal{P}_r, c \in \mathcal{C}_r, \forall r \in \mathcal{R}, t \in \mathcal{T} \tag{4}$$

## 3.3 Multi-Commodity Network Flow (MCNF) Formulation

### *3.3.1 Network flow model*

We formulate the joint optimization of patterns and headways as an MCNF problem. Figure 2 illustrates the network flow model with an example of two patterns with bi-directional consideration.

Vehicles travel from stop to stop on pattern arcs, forming a loop that includes both directions. At each origin, passengers enter a stop (of a direction) through entry arcs and are assigned a combination that defines which patterns (with corresponding headways) are available for passengers going to a particular destination. They then board the trains through a boarding arc, arrive at their destination and exit through an alighting arc.

We now introduce the variables corresponding to each arc type in the network model before presenting the MILP setting. The pattern stop sequence is represented by pattern arcs, indicating that vehicles go from stop $i \in \mathcal{S}_r$ to stop $j \in \mathcal{S}_r$ in pattern $p \in \mathcal{P}_r$. This is denoted by a binary variable $x_{tpij}$ in Eq.(5), which allows for the representation of different operational strategies such as short-turning and express service of Pattern 2 in Figure 2.[6] Operational and infrastructure constraints can be reflected by bounding some $x_{tpij} = 0$, e.g., no overtaking track or turnout available for rail.

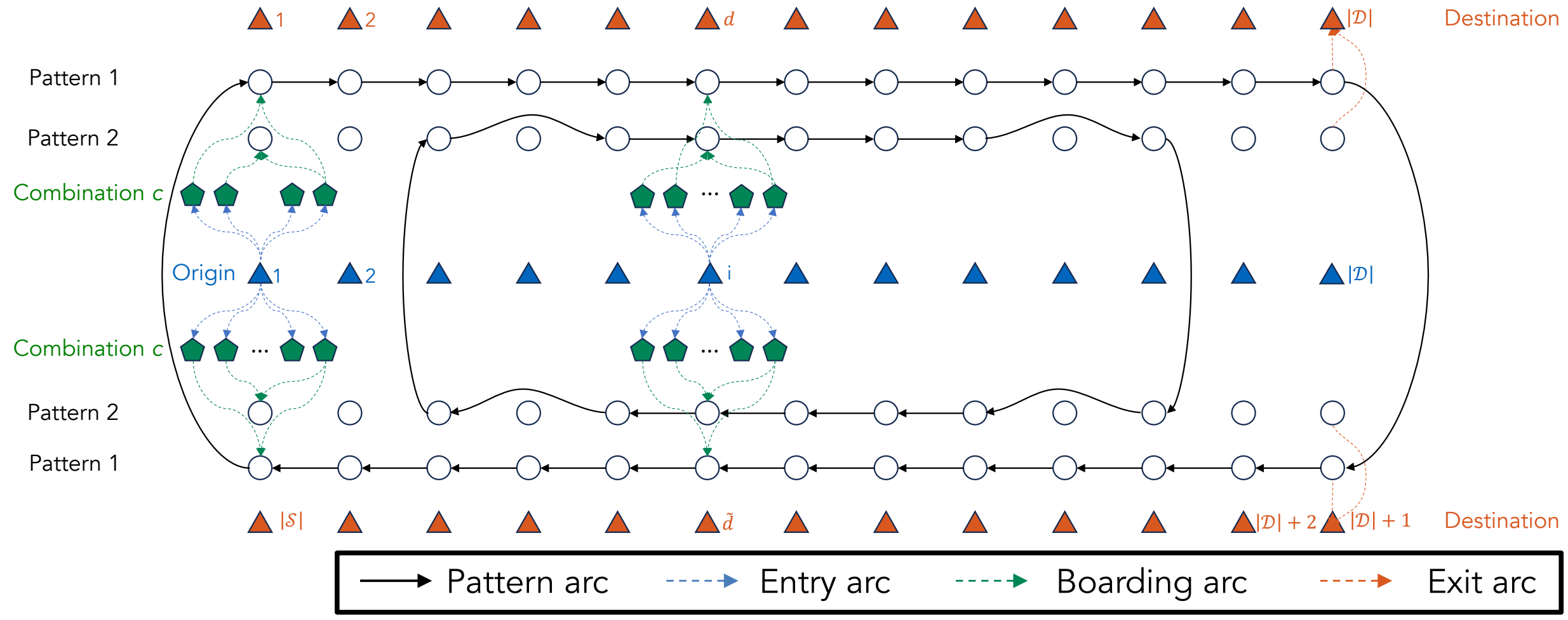

**Figure 2. Illustration of model formulation (only key nodes and links shown for clarity).**

$$x_{tpij} \begin{cases} \in \{0,1\} \text{ if } i \neq j \\ = 0 \text{ if } i = j \end{cases} \quad \forall i,j \in \mathcal{S}_r, p \in \mathcal{P}_r, r \in \mathcal{R}, t \in \mathcal{T} \tag{5}$$

Headway choices of each pattern are discretized to overcome the non-linearity between waiting times and vehicle requirements. Each pattern $p \in \mathcal{P}_r$ is assigned one headway $h \in \mathcal{H}_r$, represented by the binary variable $y_{tph}$ in Eq.(6).

$$y_{tph} = \{0,1\} \quad \forall p \in \mathcal{P}_r, h \in \mathcal{H}_r, r \in \mathcal{R}, t \in \mathcal{T} \tag{6}$$

To evaluate vehicle requirements for the stop sequence of a pattern, the binary variable $x_{tpij}$ is broken down to $x^{\eta}_{tpijh}$ in Eq.(7) to include the headway decision.

[6] Interlining is not explicitly modeled in this study but can be considered in the same framework by adjusting the network and link connection.

$$x^{\eta}_{tpijh}\begin{cases}\in\{0,1\} \text{ if } i\neq j\\ =0 \text{ if } i=j\end{cases}\quad \forall i,j\in\mathcal{S}_r, p\in\mathcal{P}_r, h\in\mathcal{H}_r, r\in\mathcal{R}, t\in\mathcal{T} \tag{7}$$

The adopted pattern-headway combination $c\in\mathcal{C}_r$ for each O-D pair $(i,d), i\in\mathcal{S}_r, d\in\mathcal{D}_r$ is represented by the binary variable $z_{tidc}$ in Eq.(8). $z_{tidc}=1$ indicates that passengers at stop $i$ going to destination $d$ take the patterns (with headways) defined in combination $c$. We note that the entry stop $i$ also indicates the direction of traveling, i.e., $i\neq\tilde{\imath}$.

$$z_{tidc}\begin{cases}\in\{0,1\} \text{ if } i\notin\{d,\tilde{d}\}\\ =0 \text{ if } i\in\{d,\tilde{d}\}\end{cases}\quad \forall i\in\mathcal{S}_r, d\in\mathcal{D}_r, c\in\mathcal{C}_r, r\in\mathcal{R}, t\in\mathcal{T} \tag{8}$$

Flows (entry $f^{\omega}_{tdic}$, boarding $f^{\alpha}_{tdicp}$, inter-stop $f^{\lambda}_{tdpij}$, ending $f^{\beta}_{tdp}$) shown in Figure 3 are continuous variables in Eq.(9)-(12). Considering passengers entering stop $i\in\mathcal{S}_r$ with destination $d\in\mathcal{D}_r$. They are first assigned to a certain pattern-headway combination $c\in\mathcal{C}_r$, i.e., the entry arc $f^{\omega}_{tdic}$. Then they board the vehicle of pattern $p\in\mathcal{P}_r$ through the boarding arc $f^{\alpha}_{tdicp}$, and travel from stop to stop through the pattern arcs $f^{\lambda}_{tdpij}$ till reaching the destination. Finally, they exit the vehicle through the exit arc $f^{\beta}_{tdp}$.

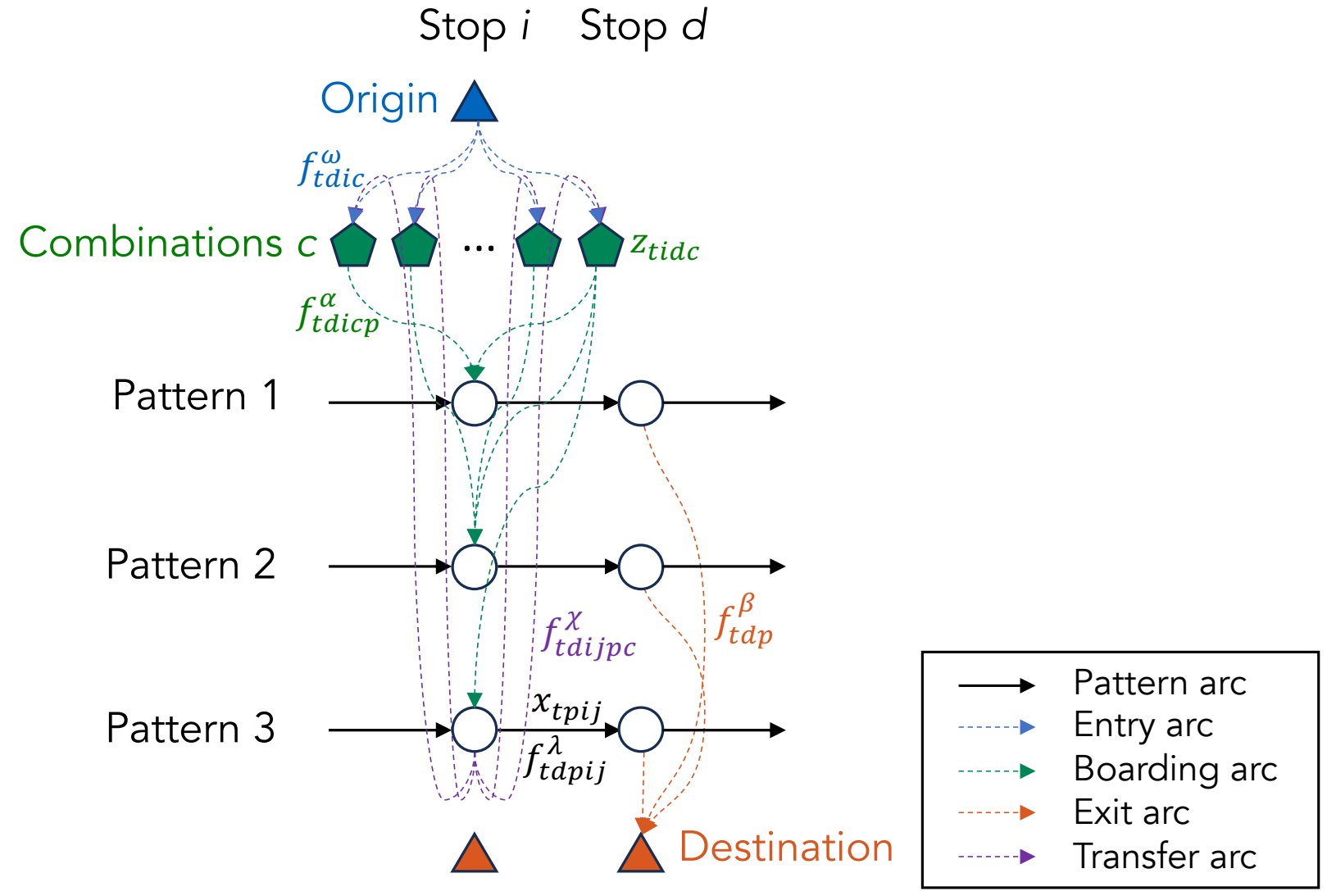

**Figure 3. Notation of flow variables.**

$$f^{\omega}_{tdic}\begin{cases}\geq 0 \text{ if } i\notin\{d,\tilde{d}\}\\ =0 \text{ if } i\in\{d,\tilde{d}\}\end{cases}\quad \forall d\in\mathcal{D}_r, i\in\mathcal{S}_r, c\in\mathcal{C}_r, r\in\mathcal{R}, t\in\mathcal{T} \tag{9}$$

$$f^{\alpha}_{tdicp}\begin{cases}\geq 0 \text{ if } i\notin\{d,\tilde{d}\}, h^{\gamma}_{cp}>0\\ =0 \text{ if } i\in\{d,\tilde{d}\} \text{ or } h^{\gamma}_{cp}=0\end{cases}\quad \forall d\in\mathcal{D}_r, i\in\mathcal{S}_r, p\in\mathcal{P}_r, c\in\mathcal{C}_r, r\in\mathcal{R}, t\in\mathcal{T} \tag{10}$$

$$f^{\lambda}_{tdpij}\begin{cases}\geq 0 \text{ if } i\notin\{d,\tilde{d}\}, i<j\\ =0 \text{ if } i\in\{d,\tilde{d}\} \text{ or } i\geq j\end{cases}\quad \forall d\in\mathcal{D}_r, i,j\in\mathcal{S}_r, p\in\mathcal{P}_r, r\in\mathcal{R}, t\in\mathcal{T} \tag{11}$$

$$f_{tjp}^{\beta} \geq 0, \forall j \in \mathcal{S}_r, p \in \mathcal{P}_r, r \in \mathcal{R}, t \in \mathcal{T} \quad (12)$$

The formulation can consider transfers across patterns. This allows better cooperation between express and local services, as some O-D pairs may not be directly connected by individual patterns. Transfer flows are represented by $f_{tdijpc}^{\chi}$ in Eq.(13) that alight a vehicle of pattern $p \in \mathcal{P}_r$ at stop $i \in \mathcal{S}_r$, to board in the same direction ($j = i$) or opposite direction ($j = \tilde{\imath}$) and join the combination $c \in \mathcal{C}_r$. These passengers would then travel with other passengers that start their journey at stop $i$ to destination $d$, therefore using the same combination $c$.

$$f_{tdijpc}^{\chi} \begin{cases} \geq 0 \text{ if } i \notin \{d, \tilde{d}\}, j \in \{i, \tilde{\imath}\} \\ = 0 \text{ if } i \in \{d, \tilde{d}\} \text{ or } j \notin \{i, \tilde{\imath}\} \end{cases} \quad \forall d \in \mathcal{D}_r, i, j \in \mathcal{S}_r, p \in \mathcal{P}_r, c \in \mathcal{C}_r, r \in \mathcal{R}, t \in \mathcal{T} \quad (13)$$

### *3.3.2 Objective*

We now introduce the MILP model. It simultaneously determines the optimal patterns and headways by setting the binary variables and evaluates the objective by solving for the minimum-cost flows through the network. The commodity of MCNF refers to the destination labels that allow checking flow balance and O-D satisfaction with much fewer flow variables. The decision variable domains have been introduced in Eq.(5)-(13) in Section 3.3.1.

The objective function in Eq.(14) aims to minimize the total weighted journey time of all passengers, formed by three main components, subject to the variable domains previously mentioned and constraints to discuss in Section 3.3.3.

The first term is the summation of riding times of all inter-stop flows from stop $i \in \mathcal{S}_r$ to stop $j \in \mathcal{S}_r$ $(i < j)$ of all destinations $d \in \mathcal{D}_r$, patterns $p \in \mathcal{P}_r$, and periods $t \in \mathcal{T}$, where $T_{ij}^{\lambda}$ is the travel time from stop $i$ to stop $j$.

The second term is the weighted waiting times of all entry flows to all combination $c \in \mathcal{C}_r$, at all stops $i \in \mathcal{S}_r$ to all destination $d \in \mathcal{D}_r$, where $T_c^{\omega}$ is the perceived headway of combination $c$ in Eq.(2) and $\gamma^{\omega}$ is a coefficient to reflect the disutility of waiting relative to riding.

The third term is the weighted transfer times of all transfer flows at station $i \in \mathcal{S}_r$, previously traveling on pattern $p \in \mathcal{P}_r$, and joining other passengers to destination $d \in \mathcal{D}_r$ who get on the station $j = i$ or $j = \tilde{\imath}$ assigned combination $c \in \mathcal{C}_r$. Each transfer costs perceived waiting time $T_c^{\omega}/2$ and transfer time $T^{\chi}$. The parameter $\gamma^{\chi}$ is a coefficient to reflect the disutility of transfers.

$$\min_{\substack{x_{tpij}, x^{\eta}_{tpijh}, y_{tph}, z_{tidc}, n^{v}_{rt} \\ f^{\omega}_{tdic}, f^{\alpha}_{tdicp}, f^{\lambda}_{tdpij}, f^{\beta}_{tjp}, f^{\chi}_{tdijpc}}} \sum_{t\in\mathcal{T}} \sum_{r\in\mathcal{R}} \sum_{d\in\mathcal{D}_r} \sum_{p\in\mathcal{P}_r} \sum_{i\in\mathcal{S}_r} \sum_{j=i+1}^{|S_r|} T^{\lambda}_{ij} f^{\lambda}_{tdpij} + \gamma^{\omega} \sum_{t\in\mathcal{T}} \sum_{r\in\mathcal{R}} \sum_{d\in\mathcal{D}_r} \sum_{i\in\mathcal{S}_r} \sum_{c\in\mathcal{C}_r} \frac{T^{\omega}_c}{2} f^{\omega}_{tdic} + \gamma^{\chi} \sum_{t\in\mathcal{T}} \sum_{r\in\mathcal{R}} \sum_{d\in\mathcal{D}_r} \sum_{i\in\mathcal{S}_r} \sum_{j\in\{i,\tilde{i}\}} \sum_{p\in\mathcal{P}_r} \sum_{c\in\mathcal{C}_r} \left(\frac{T^{\omega}_c}{2} + T^{\chi}\right) f^{\chi}_{tdijpc} \tag{14}$$

s.t. Eq.(5)-(13),(15)-(18)

### *3.3.3 Constraints*

The first batch of constraints stipulates the pattern stop sequence. The validity of a pattern is ensured by Eq.(15) that the number of incoming links to a stop equals the number of outgoing links for each pattern. Eq.(16) further restricts this number to be at most one, forming a closed loop for each pattern. $x_{tpij}$ then constrains whether flows are possible on links $(i, j)$ for pattern $p$ (only when there are services), as enforced by the big-M constraint in Eq.(17) where $M$ is a sufficiently large number[7].

$$\sum_{i\in\mathcal{S}_r} x_{tpij} = \sum_{k\in\mathcal{S}_r} x_{tpjk} \quad \forall j \in \mathcal{S}_r, p \in \mathcal{P}_r, r \in \mathcal{R}, t \in \mathcal{T} \tag{15}$$

$$0 \le \sum_{i\in\mathcal{S}_r} x_{tpij} \le 1 \quad \forall j \in \mathcal{S}_r, p \in \mathcal{P}_r, r \in \mathcal{R}, t \in \mathcal{T} \tag{16}$$

$$f^{\lambda}_{dpij} \le M x_{tpij} \quad \forall d \in \mathcal{D}_r, p \in \mathcal{P}_r, i, j \in \mathcal{S}_r, r \in \mathcal{R}, t \in \mathcal{T}, i < j \tag{17}$$

An optional constraint is bi-directional symmetry of a pattern in Eq.(18), generally more relevant to metro rail services.

$$x_{tpij} = x_{tp\tilde{j}\tilde{i}} \quad \forall i, j \in \mathcal{D}_r, p \in \mathcal{P}_r, r \in \mathcal{R}, t \in \mathcal{T}, i < j \tag{18}$$

Next, we consider the headway. Each pattern is assigned exactly one headway in Eq.(19). To break symmetries, patterns with smaller headways are assigned first in Eq.(20), which is equivalent to $h_{p_1} \le h_{p_2} \quad \forall p_1, p_2 \in \mathcal{P}_r, p_1 < p_2$

$$\sum_{h\in\mathcal{H}_r} y_{tph} = 1 \quad \forall p \in \mathcal{P}_r, r \in \mathcal{R}, t \in \mathcal{T} \tag{19}$$

$$\sum_{h'=1}^{h-1} y_{tp_1h'} \ge \sum_{h'=1}^{h-1} y_{tp_2h'} \quad \forall\, p_1, p_2 \in \mathcal{P}_r, h \in \mathcal{H}_r, r \in \mathcal{R}, t \in \mathcal{T}, p_1 < p_2 \tag{20}$$

---

[7] $M$ shall be at least as large as the maximum of $f^{\lambda}_{dpij}$.

The headway setting results also constrain $x^{\eta}_{tpijh}$, which denotes the headway and stop sequence of a pattern in Eq.(21). The variables $x^{\eta}_{tpijh}$ should also align with the pattern setting $x_{pij}$ in Eq.(22). Similar to Eq.(17), $x^{\eta}_{tpijh}$ then constrain whether flows are possible with Eq.(23) where pattern capacity is the product of vehicle capacity $B_r$ (in passengers per vehicle) and frequency $1/T^{\gamma}_{h}$.

$$x^{\eta}_{tpijh} \leq y_{tph} \quad \forall i,j \in \mathcal{S}_r, p \in \mathcal{P}_r, h \in \mathcal{H}_r, r \in \mathcal{R}, t \in \mathcal{T}, i \neq j \tag{21}$$

$$\sum_{h \in \mathcal{H}_r} x^{\eta}_{tpijh} = x_{tpij} \quad \forall i,j \in \mathcal{S}_r, p \in \mathcal{P}_r, r \in \mathcal{R}, t \in \mathcal{T}, i \neq j \tag{22}$$

$$\sum_{d \in \mathcal{D}_r} f^{\lambda}_{tdpij} \leq B_r \sum_{h \in \mathcal{H}_r} \frac{1}{T^{\gamma}_{h}} x^{\eta}_{tpijh} \quad \forall i,j \in \mathcal{S}_r, p \in \mathcal{P}_r, r \in \mathcal{R}, t \in \mathcal{T} \tag{23}$$

With $x^{\eta}_{tpijh}$, we can now evaluate the vehicle requirement based on the journey time $\sum_{i \in \mathcal{S}_r} \sum_{j \in \mathcal{S}_r} T^{\lambda}_{ij} \sum_{h \in \mathcal{H}_r} x^{\eta}_{tpijh}$ and frequency $1/T^{\gamma}_{h}$ of a pattern. In Eq.(24), the total vehicle requirement across all patterns and stops must not exceed the number of vehicles assigned for period $t$, $n^{v}_{rt}$, subject to positive fleet size for each route in Eq.(25).[8] The vehicle assignment is first constrained by the fleet size $N^{v}$ in Eq.(26), and then the sum of vehicle-hours available $N^{\tau}$ in Eq.(27).[9]

$$\sum_{p \in \mathcal{P}_r} \sum_{i \in \mathcal{S}_r} \sum_{j \in \mathcal{S}_r} T^{\lambda}_{ij} \sum_{h \in \mathcal{H}_r} \frac{1}{T^{\gamma}_{h}} x^{\eta}_{tpijh} \leq n^{v}_{rt} \quad \forall r \in \mathcal{R}, t \in \mathcal{T} \tag{24}$$

$$n^{v}_{rt} \geq 0 \quad \forall r \in \mathcal{R}, t \in \mathcal{T} \tag{25}$$

$$\sum_{r \in \mathcal{R}} n^{v}_{rt} \leq N^{v} \tag{26}$$

$$\sum_{t \in \mathcal{T}} \sum_{r \in \mathcal{R}} D_t n^{v}_{rt} \leq N^{\tau} \tag{27}$$

The third batch of constraints governs the pattern-headway combinations. Passengers at stop $i$ to destination $d$ are only assigned one combination in Eq.(28). The assignment $z_{tidc}$ is also bounded above by pattern-headway availability in Eq.(29), i.e., only combinations of selected headways of patterns serving the stop $i$ to destination $d$ can be assigned.

$$\sum_{c \in \mathcal{C}_r} z_{tidc} \leq 1 \quad \forall d \in \mathcal{D}_r, i \in \mathcal{S}_r, r \in \mathcal{R}, t \in \mathcal{T}, i \notin \{d, \tilde{d}\} \tag{28}$$

---

[8] The variable $n^{v}_{r}$ is allowed to be non-integer for cases where a train is shared across different patterns at different time. It can also be set as integers for strict operational requirement.

[9] The parameter $N^{\tau}$ reflects the total number of vehicles available, subtracted by the maintenance or workforce requirement (e.g., trains taken out of service at non-peak for maintenance, or drivers shifts).

$$z_{tidc} \leq y_{tph^{\gamma}_{cp}}, i \notin \{d, \tilde{d}\} \quad \forall i \in \mathcal{S}_r, d \in \mathcal{D}_r, c \in \mathcal{C}_r, p \in \mathcal{P}_r, r \in \mathcal{R}, t \in \mathcal{T}, h^{\gamma}_{cp} \neq 0 \tag{29}$$

The resulting $z_{tidc}$ are then upper bounds for $f^{\alpha}_{tdicp}$ in Eq.(30) to ensure no flows if a combination is not assigned. Eq.(31) further assigns flows among each pattern pairs based on frequency share rule in Eq.(4) when $z_{tidc} = 1$, where $h^{\gamma}_{cp}$ is the headway index for pattern $p$ in combination $c$.

$$f^{\alpha}_{tdicp} \leq M z_{tidc}, i \notin \{d, \tilde{d}\} \quad \forall d \in \mathcal{D}_r, i \in \mathcal{S}_r, c \in \mathcal{C}_r, p \in \mathcal{P}_r, r \in \mathcal{R}, t \in \mathcal{T} \tag{30}$$

$$\begin{gathered} -M(1 - z_{tidc}) \leq T^{\gamma}_{h^{\gamma}_{cp_1}} f^{\alpha}_{tdicp_1} - T^{\gamma}_{h^{\gamma}_{cp_2}} f^{\alpha}_{tdicp_2} \leq M(1 - z_{tidc}) \\ \forall i \in \mathcal{S}_r, d \in \mathcal{D}_r, p_1, p_2 \in \mathcal{P}_r, c \in \mathcal{C}_r, r \in \mathcal{R}, t \in \mathcal{T}, i \notin \{d, \tilde{d}\}, p_1 < p_2 \end{gathered} \tag{31}$$

The remaining constraints ensure demand satisfaction and flow balance. Eq.(32) and Eq.(33) require respectively entry and exit flows to equal O-D demand, considering flows in both directions.

$$\sum_{c \in \mathcal{C}_r} f^{\omega}_{tdoc} + \sum_{c \in \mathcal{C}_r} f^{\omega}_{td\tilde{o}c} = E_{tod} \quad \forall o, d \in \mathcal{D}_r, r \in \mathcal{R}, t \in \mathcal{T} \tag{32}$$

$$\sum_{p \in \mathcal{P}_r} f^{\beta}_{tdp} + \sum_{p \in \mathcal{P}_r} f^{\beta}_{t\tilde{d}p} = \sum_{o \in \mathcal{D}_r} E_{tod} \quad \forall d \in \mathcal{D}_r, r \in \mathcal{R}, t \in \mathcal{T} \tag{33}$$

Flow balance is handled separately for non-destination and destination stops due to the destination label. For non-destination stops, Eq.(34) ensures that the sum of entry flows and transfer flows (from both directions) equals the sum of boarding flows at each combination node. Eq.(35) ensures that sum of boarding flows and inter-stop flows (already on the vehicle) at each stop equals the sum of alighting flows and transfer flows (to both directions) at the next stop.

$$f^{\omega}_{tdic} + \sum_{p \in \mathcal{P}_r} f^{\chi}_{tdiipc} + \sum_{p \in \mathcal{P}_r} f^{\chi}_{td\tilde{i}ipc} = \sum_{p \in \mathcal{P}_r} f^{\alpha}_{tdicp} \quad \forall d, i \in \mathcal{S}_r, c \in \mathcal{C}_r, r \in \mathcal{R}, t \in \mathcal{T}, i \notin \{d, \tilde{d}\} \tag{34}$$

$$\begin{gathered} \sum_{c \in \mathcal{C}_r} f^{\alpha}_{tdjcp} + \sum_{i=0}^{j-1} f^{\lambda}_{tdpij} = \sum_{k=j+1}^{|\mathcal{S}_r|} f^{\lambda}_{tdpjk} + \sum_{c \in \mathcal{C}_r} f^{\chi}_{tdjjpc} + \sum_{c \in \mathcal{C}_r} f^{\chi}_{tdj\tilde{j}pc} \\ \forall d, j \in \mathcal{S}_r, p \in \mathcal{P}_r, r \in \mathcal{R}, t \in \mathcal{T}, j \notin \{d, \tilde{d}\} \end{gathered} \tag{35}$$

For destination stops, the sum of inter-stop flows equals the alighting flow in Eq.(36).

$$\sum_{i=0}^{j-1} f^{\lambda}_{tdpij} = f^{\beta}_{tjp}, j \in \{d, \tilde{d}\} \quad \forall d \in \mathcal{D}_r, p \in \mathcal{P}_r, j \in \mathcal{S}_r, r \in \mathcal{R}, t \in \mathcal{T} \tag{36}$$

## 4 EXPERIMENT DESIGN

We make use of a metro rail network, Chicago Transit Authority's (CTA) "L", in Chicago, IL, USA to demonstrate the use and performance of the models. The network is simplified into six lines as shown in Figure 4 (Chicago Transit Authority, 2025).[10] Stop-to-stop travel times, headway, cycle time, and number of vehicles are inferred with the current schedules. Short-turning possibilities are identified from physical rail trackwork constraints, and express services from Red-Purple dual track section. We consider one hour of continuous peak-hour operation and adapt demand data from the activity-based demand model CT-RAMP (Chicago Metropolitan Agency for Planning, 2023) and agent-based simulation tool POLARIS (Auld et al., 2016). Vehicle capacity constraints in Eq.(23) are omitted due to overall undercapacity in the system (lack of resources is more on conductors than train cars).

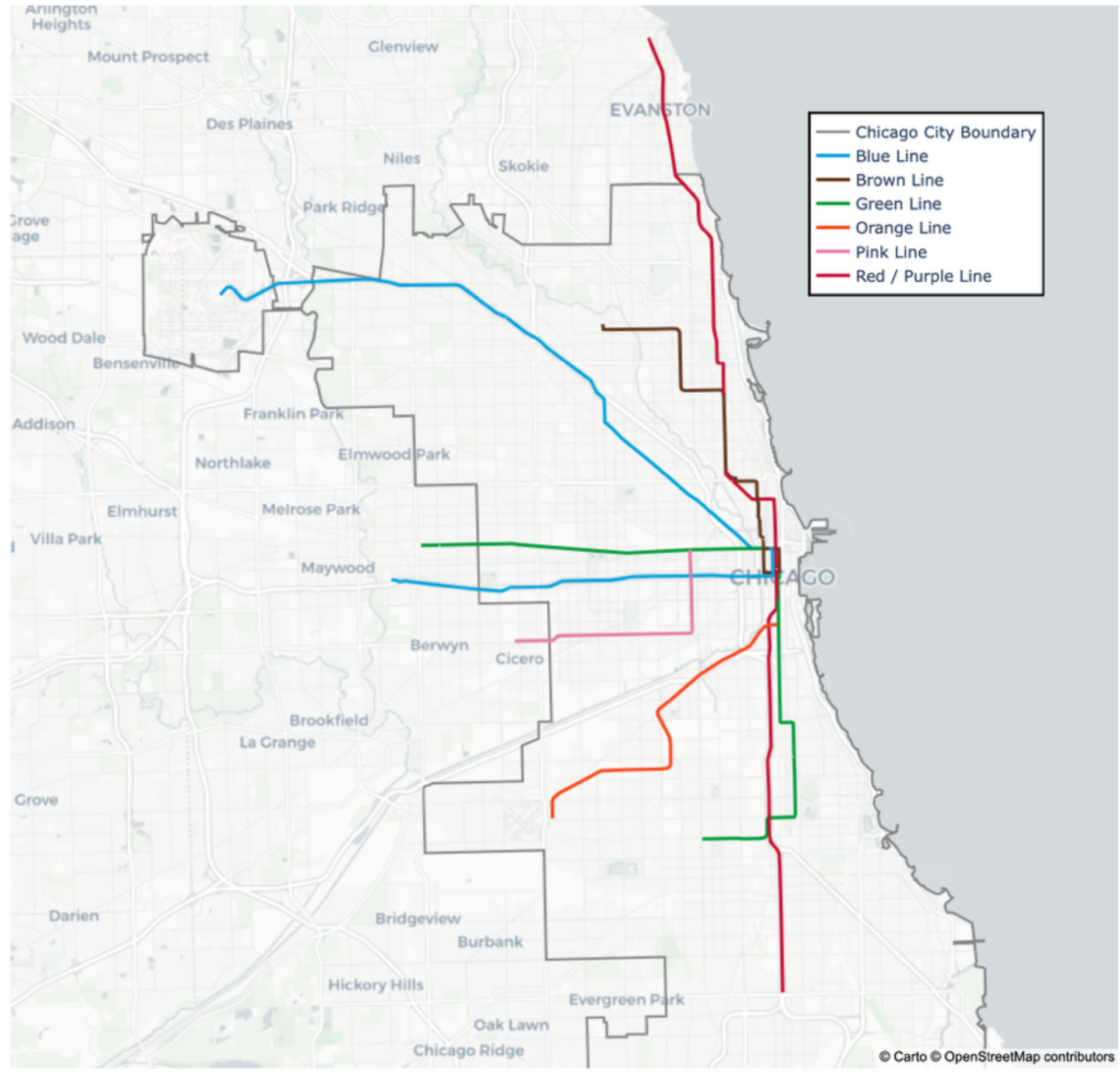

**Figure 4. Simplified network of Chicago Transit Authority's (CTA) "L" used in the experiment (Chicago Transit Authority, 2025)**

For the single-period single-route examples, we explore the following scenarios in our experiment of the Red-Purple Line, allowing transfers:

1. Baseline referencing existing Red Line and Purple Express patterns
2. Two patterns, two headway {5,7} choices per pattern
3. Three patterns, three headway {5,7,10} choices per pattern
4. Two patterns, seven headway {4,5,6,7,8,10,15} choices per pattern

---

[10] The peak-hour Purple Express and Red Line services are represented by one integrated line with express/local options. The Green Line branches are simplified to one line.

The multi-period case studies consider two design periods[11]. Single-route case study follows previous four scenarios but limits the last two scenario design to include one full pattern and no transfers within each line. Multi-route scenario, on top of the previous red line design, also optimizes train assignment across lines, short-turning, and headway of other lines (two patterns, two headway choices per pattern, one full pattern, no transfers for each line).

The optimization run was performed on a computer with Windows 11 operating system, Intel Xeon Silver 4214R 2.4GHz (12 cores), 96 GB RAM, and Gurobi 12 as the MILP solver. The parameter values are summarized in Table 2.

**Table 2. Input parameters of the models.**

| *Notation* | *Single-Period Single-Route* | *Multi-Period Single-Route* | *Multi-Period Multi-Route* |
|---|---|---|---|
| ***Sets and indices*** | | | |
| $\mathcal{D}_r$ | $\lvert\mathcal{D}_r\rvert = 43$ | | $\lvert D_r\rvert = \{33,27,28,16,22,43\}$ |
| $\mathcal{H}_r$ | $\lvert\mathcal{H}_r\rvert = 3, 4 \ or \ 8$ | | $\lvert\mathcal{H}_r\rvert = 3$ |
| $\mathcal{P}_r$ | $\lvert\mathcal{P}_r\rvert = 2 \ or \ 3$ | | $\lvert\mathcal{P}_r\rvert = 2$ |
| $\mathcal{R}$ | *{red}* | | *{blue, brown, green, orange, pink, red}* |
| $\mathcal{T}$ | $\lvert\mathcal{T}\rvert = 1$ | $\lvert\mathcal{T}\rvert = 2$ | |
| ***Parameters*** | | | |
| $D_t$ | {1} | {6,9} | |
| $E_{od}$ | Sum = 18,401 | Sum = 244,624 | Sum = 741,152 |
| $N^v$ | 40 | | 140 |
| $N^\tau$ | 40 | 540 | 2040 |
| $T_h^\gamma$ | {5,7}, {5,7,10} or {4,5,6,7,8,10,15} | | {{5,7}, {5,7}, {10,15}, {12,15}, {12,15}, {5,7}} |
| $T^\chi$ | 3 min | | |
| $\gamma^\chi$ | 2.0 | | |
| $\gamma^\omega$ | 1.5 | | |

# 5 RESULTS AND DISCUSSION

## 5.1 Single-Period Single-Route Example

We first demonstrate the single-route model with the Chicago Red-Purple Line example. Table 3 summarizes the results, showing a 0.61% reduction in objective over baseline. All solutions are solved to a 1e-4 solution gap efficiently. The two-pattern, two-headway case is solved in approximately 5 minutes, while the two more complex scenarios (around 3 and 3.5 million decision variables) are solved in less than 19 hours. This highlights the capability of the destination-labeled MCNF formulation to handle real-world pattern settings within a reasonable computational time.

[11] The two design periods are peak hours (6-9am and 4-7pm) and non-peak hours (9am-4pm and 7pm-1am).

**Table 3. Results of single-period single-route example.**

| Metric | 2 patterns, 2 headways | 2 patterns, 7 headways | 3 patterns, 3 headways |
|---|---|---|---|
| **Continuous Variables** | 843,316 | 2,267,046 | 3,284,082 |
| **Binary Variables** | 73,964 | 351,324 | 321,735 |
| **Constraints** | 1,638,176 | 3,897,186 | 5,884,864 |
| **Opt. Time (s)** | 328 | 62,212 | 67,481 |
| **Percentage Change over Baseline** | | | |
| **Objective (per hour)** | 0.00% | -0.61% | -0.61% |
| **Avg. Riding Time (min)** | 0.00% | -0.06% | -0.06% |
| **Avg. Wait Time (min)** | 0.00% | -2.21% | -2.21% |
| **Number of Transfers** | 0.00% | -92.93% | -92.93% |

Across the three optimization scenarios, the two-pattern, two-headway model obtains the existing baseline solution. The two-pattern, seven-headway and three-pattern, three-headway models obtain the same solution[12] with a 0.61% reduction in the objective function. This seemingly modest improvement translates to 37.89 passenger-hours of savings per hour, without additional operational costs. While the optimal solution maintains similar riding times, it achieves a 2.21% decrease in waiting time and eliminates most transfers.

For the resulting service patterns illustrated in Figure 5, Pattern 1 remains unchanged in the optimal solution. Meanwhile, Pattern 2 is extended to cover the southern part of the city, offering lower combined headway jointly with Pattern 1. However, Pattern 2 now runs at a higher headway (10-minute) to utilize the same total number of trains.

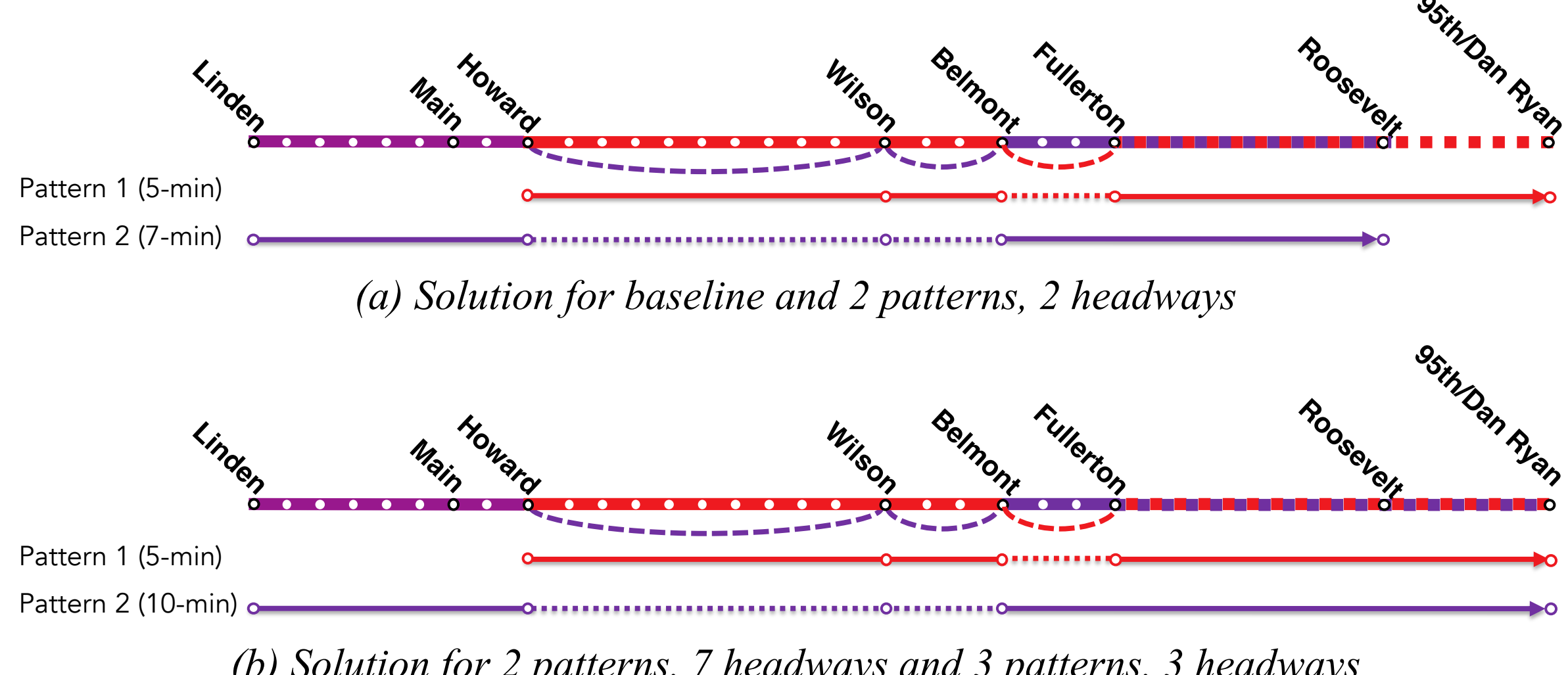

*(a) Solution for baseline and 2 patterns, 2 headways*

*(b) Solution for 2 patterns, 7 headways and 3 patterns, 3 headways*

[12] The third pattern is not used for the three-pattern case.

**Figure 5. Service pattern solutions of single-period single-route examples**

## 5.2 Multi-Period Single-Route Case Study

To demonstrate the framework's capability in optimizing route patterns and frequencies for multiple periods, we expand the example to peak and non-peak hours.

Table 4 shows the results across the four scenarios, where all solutions are solved to a 1e-4 solution gap. An objective reduction of 1.2% over baseline is shown for the 3-pattern, 3-headway case. The improvement mainly comes from the reduction in waiting times (a 9.2% decrease).

For the computational performance, the two-pattern, two-headway case is solved in around 45 minutes, 9 times the single-period case due to the solution combinations added by the two periods. The two more complex scenarios (now around 5.2 and 7.2 million decision variables) are solved in less than 9 hours.[13]

**Table 4. Results of multi-period single-route case study.**

| Metric | 2 patterns, 2 headways | 2 patterns, 7 headways | 3 patterns, 3 headways |
|---|---|---|---|
| **Continuous Variables** | 1,686,632 | 4,534,092 | 6,568,164 |
| **Binary Variables** | 147,928 | 702,648 | 643,470 |
| **Constraints** | 3,276,351 | 9,658,335 | 14,565,587 |
| **Opt. Time (s)** | 2,680 | 29,422 | 15,743 |
| **Percentage Change over Baseline** | | | |
| **Objective** | -1.20% | -0.44% | -1.23% |
| **Avg. Riding Time (min)** | -2.77% | +1.87% | +1.18% |
| **Avg. Wait Time (min)** | +3.19% | -8.16% | -9.22% |
| **Number of Transfers** | +2,299% | -100% | -100% |

The peak-hour service patterns in various scenarios are illustrated in Figure 6, where solid lines refer to all-stop service and dotted lines refer to express non-stop service. The differences lie mainly on the frequency provided to the Linden-Howard end and Roosevelt-95th/Dan Ryan end. While baseline provides 7-minute and 5-minute headways, the 3-pattern, 3-headway case provides 10-minute and 5-minute headways. The other difference is the direct/express service for different O-D trips. The last two scenarios are constrained to provide a full pattern, so the additional patterns are used to provide express service and extra frequency.

[13] The optimization times are less than the single-period case due to the restricted one full pattern and no transfers.

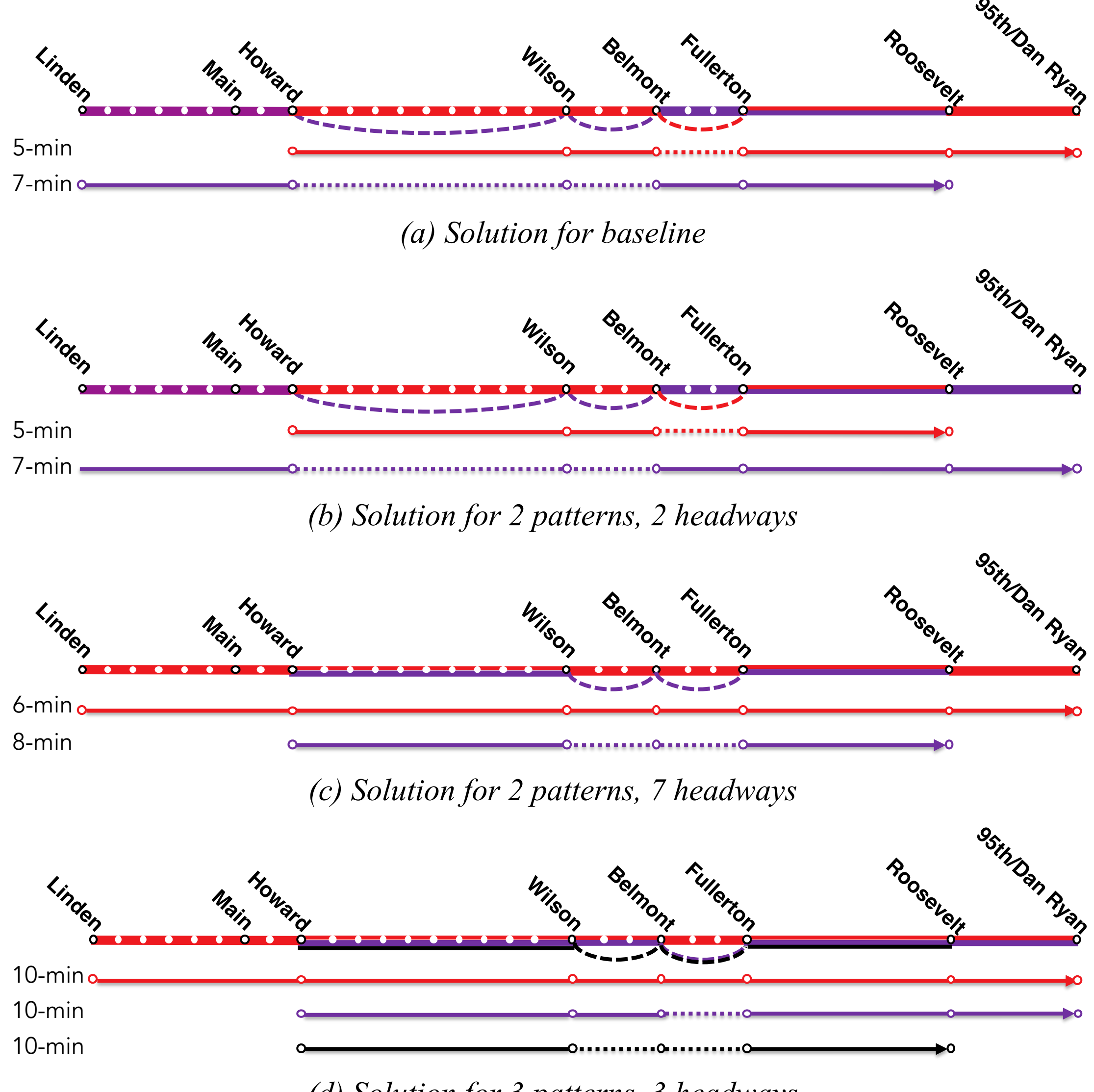

*(a) Solution for baseline*

*(b) Solution for 2 patterns, 2 headways*

*(c) Solution for 2 patterns, 7 headways*

*(d) Solution for 3 patterns, 3 headways*

**Figure 6. Peak-hour service pattern solutions of multiple-period single-route examples**

The case study demonstrates the value of the framework to optimize service pattern and headway across multiple time periods. This determines the optimal assignment of vehicles or drivers in different periods, considering the maintenance needs and demand fluctuation.

## 5.3 Multi-Period Multi-Route Case Study

On top of the previous single-route examples, we run the multi-period multi-route joint optimization model on the simplified multi-line Chicago “L” rail network. The results in Table 5 demonstrate considerable improvements in overall system performance (5.76%) with the same number of trains. Despite the model size with over four million variables, the MILP solver obtains an optimal solution quickly in slightly over four hours.

In the optimal solution, while most lines maintained their original patterns, the Green, Orange, and Pink Lines introduce two patterns each at peak hours, effectively offering smaller combined headways. Train redistribution is observed from the Blue, Brown, and Red lines to the Green, Orange, and Pink lines. This reallocation of resources allows for more frequent service on lines that benefit more.

The optimal solution results in a 5.76% objective reduction, translating to approximately 1,149 hours of travel time saved per hour of operations. This improvement is achieved through a reduction in waiting time (24.24%) with a slight increase of 0.02% in riding time. The average journey time per rider decreases by 4.17%, from 22.29 minutes to 21.36 minutes, indicating a considerable improvement in passenger experience without additional investment. This highlights the benefits of the joint optimization of patterns, headways, and fleet size across multiple lines and periods.

**Table 5. Results of multi-period multi-route case study.**

| | **Metric** | **MILP Optimization** |
|---|---|---|
| **MILP** | **Continuous Variables** | 3,683,594 |
| | **Binary Variables** | 384,352 |
| | **Constraints** | 3,064,319 |
| | **Time (s)** | 15,085 |

| | **Metric** | **Baseline** | **Optimal** | **% Change** |
|---|---|---|---|---|
| **Overall Metrics** | **Objective Function (h)** | 299,039 | 281,811 | -5.76% |
| | **Average Objective per Rider (min)** | 24.21 | 22.81 | -5.76% |
| | **Average Riding Time per Rider (min)** | 18.44 | 18.44 | 0.02% |
| | **Average Waiting Time per Rider (min)** | 3.85 | 2.91 | -24.24% |
| | **Average Journey Time per Rider (min)** | 22.29 | 21.36 | -4.17% |

| **Line** | **Metric** | **Baseline** | | **Optimal** | |
|---|---|---|---|---|---|
| | | **Peak-Hour** | **Off-Peak** | **Peak-Hour** | **Off-Peak** |
| **Blue Line** | **Headway (min)** | 5 min | 7 min | **7 min** | 7 min |
| | **Number of Trains** | 40.1 | 28.6 | **28.6** | 28.6 |
| | **Number of Patterns** | 1 | 1 | 1 | 1 |
| **Brown Line** | **Headway (min)** | 5 min | 7 min | **7 min** | 7 min |
| | **Number of Trains** | 18.8 | 13.4 | **13.4** | 13.4 |
| | **Number of Patterns** | 1 | 1 | 1 | 1 |
| **Green Line** | **Headway (min)** | 10 min | 15 min | 10, **15 min** | 15 min |
| | **Number of Trains** | 14.8 | 9.8 | **24.6** | 9.8 |
| | **Number of Patterns** | 1 | 1 | **2** | 1 |
| **Orange Line** | **Headway (min)** | 12 min | 15 min | 12, **15 min** | 15 min |
| | **Number of Trains** | 6.3 | 5 | **9.5** | 5 |
| | **Number of Patterns** | 1 | 15 min | **2** | 1 |
| **Pink Line** | **Headway (min)** | 12 min | 15 min | 12, **12 min** | 15 min |
| | **Number of Trains** | 6.8 | 5.5 | **13.6** | 5.5 |
| | **Number of Patterns** | 1 | 1 | **2** | 1 |
| **Red Line** | **Headway (min)** | 5, 7 min | 7 min | 5, 7 min | 7 min |
| | **Number of Trains** | 52.2 | 30.2 | **49.3** | 30.2 |
| | **Number of Patterns** | 2 | 1 | 2 | 1 |

# 6 CONCLUSION

## 6.1 Summary

This study develops a destination-labeled MCNF formulation for the joint optimization of transit service patterns, headways, and fleet sizes across multiple routes and periods. The model allows for flexible pattern options without relying on pre-defined candidate sets, and simultaneously considers multiple operational strategies such as route splitting, express/local services, short-turning, and deadheading. It also models bi-directional, multi-route, and multi-period operations and allows transfers across patterns in different directions.

The models are demonstrated with a case study of metro lines in Chicago, IL, USA. MILP solutions for the large-scale problem are obtained efficiently. For the single-route example, the model achieves a 0.61% reduction in total weighted journey time without additional operational costs. This seemingly modest improvement translates to 37.89 hours of passenger time saved per hour of operation. In the multi-period multi-route scenario involving six lines, the joint optimization results in a more substantial 5.76% reduction in total weighted journey time, equivalent to approximately 1,149 hours saved per hour of operations. This is achieved primarily through a 24.24% decrease in waiting times, without additional operational cost.

## 6.2 Applications and Implications

The proposed MCNF formulation can be applied for transit planning and operations in different line-based modes, such as metro rail, light rail, and buses. The efficient joint optimization of patterns, headways, and fleet sizes across multiple routes provides transit agencies with a tool for comprehensive service design and resource allocation in city-scale networks. Different operational strategies can be considered for various routes simultaneously subject to fleet size and physical constraints.

In the broader multimodal transit network design efforts (Ng et al., 2024), the pattern optimization in this study considers routes in a transit network and generates patterns for transit operators. These are inputs to the next-step frequency and shared autonomous vehicle fleet size setting (Pinto et al., 2020). This helps to create more efficient and integrated transportation systems, improving service quality amid limited transit budgets.

### 6.3 Limitations and Future Research

While the study presents significant advancements, there are several limitations and areas for future research:

1. Computational requirements: The model's complexity increases rapidly with the number of stops, patterns, and headway options. Current computation limitations make multi-period solutions difficult with route flexibility and transfer options. Future work could compare and accelerate the solutions with other programming and heuristic approaches in larger networks and more granular headway choices.
2. Inelastic demand: The current model assumes fixed demand. Incorporating elastic demand would allow for a more realistic representation of how service improvements attract additional ridership. This could be addressed through an iterative process between service optimization and demand forecasting.
3. Homogeneous train configuration: Different train lines may use various number of train cars for capacity, cost, and operational consideration. Future study can optimize the number of train cars as well subject to the operation scenario.

In conclusion, this study presents an efficient tool for transit agencies to optimize their service patterns, headways, and fleet allocation. The demonstrated ability to achieve considerable efficiency gains without additional operational costs underscores the potential of this approach to improve transit service quality and resource utilization.